\documentstyle[12pt]{ioplppt}

\begin{document}
\jl{2}
\parindent0cm

\title{Coupling constants for a degenerate Fermi gas confined to
a quasi one-dimensional harmonic trap}[Coupling constants 
for fermions in  one dimension]

\author{F Gleisberg and W Wonneberger}

\address{Abteilung Mathematische Physik, Universit{\"a}t Ulm, D-89069 Ulm, Germany}

\begin{abstract}
A theory for the coupling constants between spin 
polarized degenerate fermionic atoms is developed for the 
case of atoms confined to a quasi one-dimensional harmonic trap. 
Exploiting the presence of the Fermi edge and the large number of fermions it is shown 
that the resulting effective one-dimensional interaction can be parameterized by three 
sets of interaction 
coefficients, two for forward scattering and one for backward scattering. In the case of 
identical fermions, backscattering dominates because the contact part of the 
effective one-dimensional interaction must be subtracted. Analytic expressions for the 
interaction coefficients for the effective contact interaction, which is relevant for 
the inter-component interactions, and for the "p-wave" interaction appearing in next 
order, are given. As an example, we calculate and discuss in detail the 
effective coupling constants for the marginally long ranged dipole--dipole interaction. 
\end{abstract}


\maketitle

\section{Introduction}

The achievement of Bose-Einstein condensation in dilute ultracold gases \cite{ADB95} 
also renewed the theoretical interest in fermionic many body systems \cite{BDL98,BB98,ZGOM00,Yoshi},
e.g. their superfluid properties \cite{HFSt97,BP98,HSt99,C99,Shl02}.
Experimental successes in obtaining degeneracy in Fermi 
vapours \cite{DMJ99,OHara00,SFCCKMS01,Tru01,SK01} further stimulated the interest in confined 
ultracold Fermi gases. Recently, successful production of large samples of fermionic lithium ($10^6$ - $10^7$ atoms)
with long lifetimes ($~10^2$ sec.) was reported \cite{Hadz02,Hadz03}.

The present situation in this field is characterized by intensive search
for a verification of the presence of a superfluid BCS-like phase.
Strong attractive interaction between the fermions is needed for a
transition temperature not too far below the temperature of
Fermi degeneracy. It is not only the absence of s-wave scattering between
identical fermions which causes problems, even the intrinsic
interaction between non-identical fermions is generally
too weak.
Strongly interacting degenerate Fermi gases now can be
produced either
by mixing bosons with fermions \cite{Heiselberg00,Roati02} or by magnetically inducing
a Feshbach resonance in a two-species mixture of fermions
\cite{Timmermans01,Holland01,Ohashi02,OHara02}.

Here we study the interaction of both identical 
and non-identical fermions in
a one-dimensional harmonic trap where no superfluid transition 
can occur at finite temperatures.
The interactions considered  in the present paper are limited to scattering in a
single channel when energy is exchanged between the scattering
particles and the harmonic trap. Moreover,
a large interaction energy would violate the criterion for
a one-dimensional regime as is discussed in Section 2.

Using microtrap technology \cite{VFP98,FGZ98,DCS99,RHH99,Ott01}, it will become
possible in the near future to produce a neutral ultracold  quasi one-dimensional  
Fermi gas in a highly elongated trap. 

The confinement of a trapped ultracold gas can be realized by a harmonic external 
potential.
Theoretical work on interacting 
fermions in a one-dimensional  harmonic trap has been published in \cite{WW01_2,Recati}. It is based on
concepts of Luttinger liquid theory \cite{Haldane} (for reviews see \cite{E79,V95,Sch95}).
Detailed knowledge of the coupling constants will be necessary for understanding the 
ground state properties of this novel confined one-dimensional Fermi gas. The coupling constants
will decide which type of fluctuations are dominant, density wave, component wave, or 
pairing fluctuations. 

Solvable models require specific forms of the interaction
coefficients, especially the exclusion of momentum
non-conserving collision events in the harmonic trap (which
evidently are present here). We study the coefficients of the pair interaction
in order to find out to what extent physical interactions have at
least approximately the behaviour required by the exactly 
solvable models.

We construct an effective one-dimensional potential from a physical interaction potential using
the simplest conceivable realization of the projection from three dimensions to one dimension.
We set up the effective interaction operator 
by taking into account only interaction coefficients of the transverse ground state
and study the effects of the one-dimensional geometry on the form of the interaction
coefficients.

In many cases, identical spin polarized fermions experience only 
a weak residual interaction because s-wave scattering is forbidden. 
This restriction does not hold for a two-component system of spin 
polarized fermions where significant 
interactions between the components are possible. For instance, the  
dipole--dipole interaction \cite{GER01} can become relevant, especially in the case 
of polar molecules \cite{BBC00}.

Postulating the existence of an effective local one-dimensional potential with a Fourier transform,
we demonstrate that three main scattering processes dominate the interaction coefficients
in one dimension. 
They are identified as two forms of forward and one form of backward scattering. They 
determine the coupling constants in theories of the interacting one-dimensional 
degenerate Fermi gas.

Recently, a two-component Fermi gas has been realized \cite{JinPappDeMarco}
with different hyperfine states of $^{40} \rm K$. For simplicity, however, we
assume the same longitudinal trapping frequency
for both components.

The paper is organized as follows. Section 2 develops the theory for the interaction 
coefficients. It includes a discussion how the one-dimensional regime may be reached.
In Section 3 we evaluate the interaction coefficients
for an arbitrary interaction of finite range in terms of
the zeroth and second moments of the effective one-dimensional interaction
potential. In Section 4 we present the results
for the marginally long-ranged dipole-dipole
interaction. In both cases we assess the relevance
of our findings for one- and two-component
Fermi vapours in one dimension.

\section{Theory}

Starting from the case of identical atoms in the same hyperfine state,
the pair interaction operator ${\hat V}_3$ for a many particle system in a trap is

\begin{equation}\label{1.1a}
{\hat V}_3 = \frac{1}{2}\sum\, V(\mbox{\boldmath$m$},\mbox{\boldmath$p$};\mbox{\boldmath$q$},\mbox{\boldmath$n$})\,
{\hat c}_{\mbox{\boldmath$m$}}^+{\hat c}_{\mbox{\boldmath$q$}}^{\,}\,\,{\hat c}_{\mbox{\boldmath$p$}}^+ {\hat c}_{\mbox{\boldmath$n$}}^{\,},
\end{equation}

with interaction coefficients

\begin{equation}\label{1.2a}
V(\mbox{\boldmath$m$},\mbox{\boldmath$p$};\mbox{\boldmath$q$},\mbox{\boldmath$n$})=\int \d^3 x\,\d^ 3x'\,\psi^*_{\mbox{\boldmath$m$}}({\bf x})
\psi^*_{\mbox{\boldmath$p$}}({\bf x}')
V_3({\bf x}-{\bf x}')\psi _{\mbox{\boldmath$q$}}({\bf x})\psi _{\mbox{\boldmath$n$}}({\bf x}').
\end{equation}

$V_3({\bf x}-{\bf x}')$ denotes the two-particle interaction potential while
the quantum numbers $m_x,m_y,m_z$ etc., abbreviated by $\mbox{\boldmath$m$}$, characterize the
single particle bound state in the potential of the trap.

In our case,  

\begin{equation}\label{1.3a}
V({\bf x}) = \frac{1}{2} m_{\rm A} \omega ^2 _{\ell} \,z^2 +\frac{1}{2} m_{\rm A} \omega _\perp^2 \,
(x^2+y^2)
\end{equation}

is the potential of the highly elongated (trap frequencies $\omega _\perp \gg \omega _\ell$)
axially symmetric harmonic trap. The atomic mass is denoted by $m_{\rm A}$.
The non-interacting system has single particle levels

 \begin{equation}\label{1.4a}
 \hbar \omega _{\mbox{\boldmath$n$}} = \hbar \omega _{\ell} \left(n_z+\frac{1}{2}\right)+\hbar\omega _\perp
 (n_x+n_y+1),\qquad n_x,n_y,n_z=0,1,\ldots
 \end{equation}

and wave functions which are products of the 
wave functions of the one-dimensional harmonic oscillator, e.g., for the
coordinate $z$ in the elongated direction of the trap ($n=n_z$)
 
 \begin{equation}\label{1.4}
 \psi _n(z) = \left(\frac{\alpha}{2^n n!\pi^{1/2}}\right)^{1/2} 
 \,\exp(-\alpha^2 z^2/2)\,{\rm H}_n (\alpha z).
 \end{equation}

 ${\rm H}_n$ denotes a Hermite polynomial.
 Intrinsic length scales of the trap are $\ell = \alpha ^{-1}$ 
 with  $\alpha^2=m_{\rm A} \omega _\ell/\hbar$ and $\ell _\perp=\alpha _\perp^{-1}$ with
$\alpha _\perp^2=m_{\rm A} \omega _\perp/\hbar$, the spatial
extensions of the longitudinal and transverse ground states, respectively. 

We consider a system of $1 \ll N<\omega _\perp/\omega _\ell$ fermionic atoms. At $T=0$, 
they fill
the lowest $N$ states well below the first transverse excited state. Note that
each single particle level 
is occupied by at most one spin-polarized fermion. Then the unperturbed Fermi energy is

 \begin{equation}\label{1.3}
 \epsilon _{\rm F} = \hbar \omega _\ell \left(N-\frac{1}{2}\right).
 \end{equation}

The transverse ground state energy $\hbar \omega _\perp$ is taken from now on as zero of energy.

Neglect of all terms with transverse quantum numbers not equal to zero defines our quasi
one-dimensional model. The validity of this approximation is discussed below.
From now on we write $m$ for $\mbox{\boldmath$m$}=(0,0,m_z)$ etc.
and use an effective one-dimensional potential $V_{\rm 1eff}$ which results from averaging the
physical potential $V_3$ over the transverse ground state

\begin{equation}\label{1.7a}
V_{\rm 1eff}(z-z')=\int \d^2x\, \d^2x'\,\psi _{\perp 0}(x)^2 \psi _{\perp 0}(y)^2 \psi _{\perp 0}
(x')^2 \psi _{\perp 0}(y')^2 V_3({\bf x}-{\bf x}').
\end{equation}

For example, a three-dimensional contact potential $V_3({\bf x})=g\,\delta^{(3)}({\bf x})$ 
(which contributes only for bosons or non-identical fermions to ${\hat V}_3$) 
gives 

\begin{equation}\label{1.8a}
V_{\rm 1eff}(z)= g\frac{\alpha^2_\perp}{2\pi}\delta(z).
\end{equation}

Instead of computing $V(m,p;q,n)$ from the real space potential equation (\ref{1.7a})
we use in our calculations below a Fourier version of equation (\ref{1.2a}) which reads

\begin{eqnarray}\label{1.5}
\fl V(m,p;q,n)=\cos(\pi(|q-m|-|n-p|)/2)
\left(2\,\frac{{\rm min}(m,q)! \,{\rm \min}(n,p)!}{{\rm max}(m,q)! \,{\rm max}(n,p)!}\right)^{1/2}
\\ \nonumber
\times\frac{\alpha}{2 \pi}\int _0^\infty \d v\,v^{(|q-m|+|n-p|-1)/2}\,e^{-v}
\tilde{V}_{\rm 1eff}(k_\|^2=2 \alpha^2 v)\,{\rm L}_{{\rm min}(q,m)}^{(|q-m|)}(v)
{\rm L}_{{\rm  min}(n,p)}^{(|n-p|)}(v), 
\end{eqnarray} 

provided the Fourier transform ${\tilde V}_{\rm 1eff}(k_\|)$ of $V_{\rm 1eff}(z)$ 
\[
\tilde{V}_{\rm 1eff}(k_\|)=\int \d z \exp (\i k_\| z)
V_{\rm 1eff}(z)
\]

exists.
The $z$--component of the Fourier variable $\bf k$ is denoted by $k_\|$.
${\rm L}_i^{(j)}$ denotes a Laguerre polynomial.

We generalize to the case of two components and
discuss the conditions under which the one-dimensional model defined above is valid.
First of all we exclude thermal excitations of 
the transverse levels by demanding
$k_B T \ll \hbar \omega_\perp$.
Consider now $2N$ fermionic atoms 
in an elongated harmonic trap
which are distributed equally into 2 hyperfine states.
Atoms of different components 
interact through a
three-dimensional contact potential $g\,\delta^{(3)}({\bf x}-{\bf x}')$.
Interaction between atoms of the same component is much weaker 
and is therefore
neglected here.
The relation $N<\omega_\perp/\omega_\ell$ guarantees 
that all the atoms can be accomodated in the non-interacting ground state.
We use the mean field estimate
$n g$ for the interaction energy per atom denoting the
density in three dimensions of each hyperfine component by $n$. Following D. Gangardt and
and G. Shlyapnikov \cite{GShly}, this energy must 
be much smaller than the transverse excitation energy $\hbar\,\omega_\perp$
to be in the one-dimensional regime. 
Hereby it is assumed that the width of the transverse 
ground state of the trap greatly exceeds the
three-dimensional scattering length $a=mg/4\pi\hbar^2$ for the scattering of fermions
in different hyperfine states \cite{O98}.
$N$ atoms in the transverse ground state of the trap
occupy a volume of $\sim 2\pi\,\alpha^{-1}\,\alpha_\perp^{-2}\,\sqrt{2N}$
and the above criterion may be written as
\begin{equation}\label{Criterion} 
a\, \alpha \, \sqrt{2N}\,\, \ll \, \,1.
\end{equation} 
Note that the
aspect ratio $\left(\omega_\ell/\omega_\perp\right)^{1/2}$ does not enter
the criterion (\ref{Criterion}) explicitely. It is not only a sufficiently large ratio of
$\omega_\perp/\omega_\ell$ but the reduction of $N$ which
leads into the one-dimensional regime provided the Fermi level is always below
$\hbar \omega_\perp$.

As an example we consider fermionic potassium atoms in 2 different
hyperfine states in an elongated harmonic trap with typical parameters:
$\omega_\ell/\omega_\perp=10^{-4}$, 
$\alpha=10^3 \,\,\mbox{cm}^{-1}$, and $a=8\,\,\mbox{nm}$.
The transverse ground state of the trap be half filled by each component.
Then the ratio between the energies of interaction and transverse
excitation is as small as $~ 0.05$. In what follows 
we assume that the above criterion is always fulfilled.

We continue to discuss the interaction coefficients $V(m,p;q,n)$.
From equation (\ref{1.7a}) the following features may be deduced:
If the atoms interact by 
the marginally long ranged dipole-dipole
interaction which has no intrinsic range, the effective one-dimensional potential
has a finite range
$d_{\rm 1D} \sim \ell_\perp$. This is shown in detail in
Section 4.
 We note that a similar calculation
may be performed for the
Coulomb interaction which has an infinite range and results in
$d_{\rm 1D}\sim 3\ell_\perp$. 
If instead the interatomic potential has a finite range $d_{\rm 3D}$ as e.g.
the van der Waals interaction then we find $d_{\rm 1D} \sim d_{\rm 3D}$.

In addition, the integrand
in  equation (\ref{1.5}) contains further functions including Laguerre 
polynomials. This weighting provides a soft momentum cut-off at $k >2k_{\rm F}$ or $v >
4N$, provided we consider the physically relevant case $m \approx n \approx p \approx q
\approx N \gg 1$, but $|m-n|\ll N$, $|m-q|\ll N$, etc.. It is here, where the fermionic 
nature of the quantum gas becomes relevant.

There are 24 coupling coefficients $V(m,p;q,n)$ for any set of four distinct integers. Due 
to the 
symmetries $(m \leftrightarrow q)$, $(n \leftrightarrow p)$, and $([m,q]\leftrightarrow 
[n,p]) $ only three of them can be different.

It is one of the aims of the present work to study how far
atoms confined to a one-dimensional harmonic trap and interacting by physical potentials 
have interaction coefficients $V(m,p;q,n)$ which behave as

\begin{eqnarray}\label{1.6}
V (m,p;q,n) \approx V _{\rm a} \, \delta _{m-q, n-p}+V _{\rm b} \, \delta _{q-m, n-p}+
V _{\rm c} \, \delta _{m+q, n+p}.
\end{eqnarray}

For such an interaction an exact solution of the model 
was found in previous work \cite{WW01_2}.
Fermions in a box have exactly the required behaviour as is shown
in the appendix. This is a consequence of local translational invariance inside the box.
In the one-dimensional harmonic trap, however, the situation
is different. Only in a region around the center of the trap, say at $|z| \le \delta$, the atoms
nearly feel no external force.   
Here the single particle states are superpositions of plane wave states
$\exp(\i k_nz)$ with $k_n=\pm\alpha\sqrt{2n-1}$. Because the relevant states
are near the Fermi edge, we have $|k_n|\approx k_{\rm F}$.
According to equation  (\ref{1.1a}), 
incoming states $\{n,q\}$ are transformed into
outgoing states $\{p,m\}$ in the collision process. Denoting a state
with $k_n \approx -k_{\rm F}$ by $(-n)$, three types of 
collision processes may
be discriminated 

\begin{equation}\label{Table}
\begin{array}{llcl}
\mbox{a-type:  } & \{n,q \} & \rightarrow & \{p,m\},\\
\mbox{b-type:  } & \{n,-q\} & \rightarrow & \{p,-m\},\\ 
\mbox{c-type: } & \{n,-q\} & \rightarrow & \{-p,m \}.
\end{array} 
\end{equation}
If the atoms move in the same direction, only forward scattering
($\Delta p \ll \hbar k_{\rm F}$, type a) may occur.
If the atoms move in opposite directions, forward scattering 
($\Delta p \ll \hbar k_{\rm F}$, type b) as
well as backward scattering ($\Delta p \approx 2\hbar k_{\rm F}$, type c)
are possible. 
The first two cases were considered in the first two papers of \cite{WW01_2}. The last one requires an 
extension of the bosonization method, which is the aim of the third paper \cite{WW01_2}.
The couplings $V_{\rm a}$, $V_{\rm b}$, and $V_{\rm c}$ are the analogues of the Luttinger model couplings 
$g_4$, $g_2$, and $g_1$, respectively \cite{E79,V95,Sch95}. 

Because the quantum numbers $m,p,q,n$ denote energies
in units of $\hbar \omega_\ell$, momentum conservation
can only be approximately fulfilled. If e.g. in the
c-type collision process $m+q-p-n=0$ then, linearizing 
with respect to $m-N$ etc. gives
\begin{equation}\label{momenta}
|k_m| + |k_q| - |k_p| - |k_n| = O(N^{-1}),
\end{equation}
and one would conclude that the coefficients
$V_{\rm a,b,c}$ would dominate more clearly with increasing $N$.

On the other hand, the larger $N$, the larger is the fraction of atoms
in the regions $|z| >  \delta$ where momentum is not conserved (recall
that the number of atoms in the central region
$|z| \le \delta$ is only $\propto \sqrt{N}$)
and this would enhance the momentum non-conserving coefficients.

From the discussion below it will turn out that for short range
interactions (approximated by a contact potential) coefficients
of type a,b, or c 
dominate over the coefficients with indices which do not 
fulfill one of the three momentum conservation relations above.
Moreover, we will show that backscattering is strongly suppressed 
for interactions with a range of more than the interparticle
separation.

\section{General Discussion}

Before going to specific interactions, we address some general questions related
to the above results.

Interactions produced by a one-dimensional contact potential will henceforth be called "s-wave".
Such a contact potential produces interaction coefficients, which are fully symmetric in
all their arguments. Therefore they do not contribute to the interaction between identical
fermions. 
We demonstrate how a contact contribution is subtracted from
the interaction coefficient

\begin{equation}\label{Gen1}
V(m,p;q,n)=\int \d z\,\d z'\,\psi _m(z)\psi _q(z)\,V (z-z')\,
\psi _p(z') \psi _n(z').
\end{equation}

The one-dimensional potential $V(z)$ could, for instance, be an effective
one-dimensional potential calculated from a physical potential via equation (\ref{1.7a}).
If $V (z)$ decays faster than the product $
\psi _p(z') \psi _n(z')$ oscillates
(for $p \approx n \approx N$ the wavelengths are
about $\pi/k_{\rm F} \approx \pi \ell _\perp$) one can expand
$\psi _{p}(z')\psi _n(z')$ at $z'=z$ up to second order and gets

\begin{eqnarray}
V(m,p;q,n) & = & \langle V \rangle \, \int \d z\,\psi _m(z)\psi _q(z)\,
\psi _p(z) \psi _n(z) \nonumber\\[3mm]
 &   &\,\,  + \,\frac{1}{2}\,\langle z^2\,V \rangle
\, \int \d z\,\psi _m(z)\psi _q(z)\,
\frac{\d^2}{\d z^2}\,\left[\psi _p(z) \psi _n(z)\right] + \ldots \nonumber\\[3mm]\label{Gen2}
 & = & V^{({\rm s})}(m,p;q,n) + V^{({\rm p})}(m,p;q,n) + \ldots
\end{eqnarray}

We call the first term in equation (\ref{Gen2}), which corresponds to a contact
interaction, $V^{({\rm s})}$ (''s--wave'') and
the second term $V^{({\rm p})}$ (''p--wave'') contribution to the interaction
coefficient.
Higher order "partial waves" can be defined analogously.

A sufficient condition 
for the validity of the expansion is $ 2 k_{\rm F} d_{\rm 1D} \ll 1$. 
Here we use the notation $\langle V\rangle = \int dz\, V(z)$
and assume that the first moment $\langle z\,V\rangle$ of $V $ vanishes
because $V $ is an even function.
The moments of $V $ may be expressed in terms of
the Fourier transform ${\tilde V} (k)$ of the potential and its second derivative 
$\tilde{V}''(k)$ at $k=0$

\begin{equation}
\langle V \rangle =\int \d z\, V (z) = {\tilde V}  (k=0),
\end{equation}
\begin{equation}
\langle z^2\,V \rangle =\int \d z\,z^2\, V (z) = - {\tilde V} ^{''} (k=0),
\end{equation}

provided the integrals converge.
The Fourier variable $k$ conjugate to $z$ in this case corresponds to $k_\|$
as explained above.
   
We note that the expansion equation (\ref{Gen2}) is equivalent to 
inserting the approximation

\begin{equation}\label{Gen5}
V (z) \approx \langle V \rangle \,\delta(z)+ \frac{1}{2}\,
\langle z^2 V \rangle \,\frac{\d^2}{\d z^2}\delta(z)+ \ldots
\end{equation}

into equation (\ref{Gen1}).

In a contact interaction, the momentum transfer is uniform: Both forward scattering
($k \ll k_{\rm F}$) as well as backward scattering ($k \approx 2k_{\rm F}$) are equally present in the
potential and hence contribute equally to the first term in equation (\ref{Gen2}). 
In "p-wave" scattering, however, the second derivative leads to a 
suppression of the forward contribution. 

In the Fourier 
domain, equation (\ref{Gen5}) reads

\begin{equation}\label{2.4}
\tilde{V}(k)=
 \tilde{V} (k=0)+\frac{1}{2}\,\tilde{V}''(k=0)\,k^2 + O(k^4),
\end{equation}

provided ${\tilde V} (k=0)$ and $\tilde{V}''(k=0)$ exist.
Again, the first term is called ''s--wave'' and the second ''p--wave''.

The dominant interaction between different components in a mixture of Fermi gases 
with short range interaction
is the ''s-wave'' part. It is characterized by interaction coefficients, which are 
fully symmetric in its arguments.

In this case, inserting only the ''s--wave'' part 
$\tilde{V}(k=0)$ of equation (\ref{2.4}) into equation (\ref{1.5}), the $v$-integration can be performed 
exactly, leading to the result: 

\begin{eqnarray}\label{2.7}
\fl V^{({\rm s})}(m,p;q,n) =  \alpha \langle V \rangle \,
\frac{\cos(\pi (p+q-m-n)/2) }{\sqrt{2}\pi}
\left(\frac{\Gamma(m+1)}{\Gamma(p+1)\Gamma(q+1)\Gamma(n+1)}\right)^{1/2}\nonumber\\[4mm]
\lo \times \frac{\Gamma(a_1)\Gamma(a_2)}
{\Gamma(a_3)\Gamma(m-q+1)}
\, \left._3F_2\left(-q,a_4,a_2;  
 a_5, m-q+1 ; 1\right)\right.
\end{eqnarray}
with the abbreviations
$a_1=(p+q+n-m+1)/2$,
$a_2=(m+p-q-n+1)/2$,
$a_3=(p+q-m-n+1)/2$,
$a_4=(m+n-p-q+1)/2$,
$a_5=(m-n-p-q+1)/2$.

The generalized hypergeometric function  $_3F_2$ reduces here to a finite series \cite{Pr}.
 
The coefficients $V^{({\rm s})}(m,p;q,n)$ 
have arguments $m,p,q,n \approx N$ while the differences of the
arguments are much smaller than $N$ as was explained above. In this case,
the relative variation of the arguments of the functions 
which determine $V^{({\rm s})}(m,p;q,n)$ according to equation (\ref{2.7}) 
is small with respect to a 
variation of $N$ but of order unity with respect to variations of
the differences of the arguments. 
Numerical examination
of equation (\ref{2.7})
shows that $V^{({\rm s})}(m,p;q,n)$ behaves indeed in this way. 

Following equation (\ref{1.6}), we use the notation $V_{\rm a}$ for $V(m,p;q,n)$ with $m+p=q+n$.
$V_{\rm b}$ and $V_{\rm c}$ are defined analogously. Specifically for the coefficients $V_{\rm a}$,
$V_{\rm b}$, and $V_{\rm c}$, we find from equation (\ref{2.7})
\begin{equation}\label{Exponent}
V^{({\rm s})}_{{\rm a,b,c}} (m,p;q,n \approx N) \propto N^{-0.3}.
\end{equation}

For the model interaction in the appendix (see equation (\ref{A.6}) below)
we have instead $V_{\rm a,b,c} \propto N^{-1/2}$ (note $L\approx L_{\rm F}$) 
in accordance with an argument given in \cite{Recati}.
In contrast, the momentum non-conserving
coefficients are almost independent of $N$. Concerning
the neglect of all background coefficients,
as in the Tomonaga-Luttinger model with harmonic
confinement, we conclude from the present work
that this model may be applicable to non-identical fermions
if the number of atoms in the trap is limited \cite{model}.

In figure  1, we show corresponding results.
We have chosen the form of a histogram since 
the coefficients $V(m,p;q,n)$ with an odd sum
of arguments vanish.

The ''p-wave'' interaction coefficients
are obtained by inserting 
only the ''p--wave'' part 
$\frac{1}{2}\tilde{V}''(k=0)\, k^2$ of equation (\ref{2.4}) 
into equation (\ref{1.5}). They can also be given in closed form: 

\begin{eqnarray}\label{2.8}
\fl V^{({\rm p})}(m,p;q,n) = - \alpha^3\,\langle z^2 V\rangle \,
\frac{\cos (\pi (p+q-m-n)/2) }{\sqrt{2}\,\,\pi}
\, \left(\frac{\Gamma(m+1)}{\Gamma(p+1)\Gamma(q+1)\Gamma(n+1)}\right)^{1/2}
\nonumber\\[4mm]
\lo \times\, \frac{\Gamma(b_1)\Gamma(b_2)}
{\Gamma(b_3)\Gamma(m-q+1)}
\,\left._3F_2\left(-q, b_4, b_2; b_5, m-q+1; 1 \right)\right.
\end{eqnarray}
with the abbreviations
$b_1=(p+q+n-m-1)/2$,
$b_2=(m+p-q-n+3)/2$,
$b_3=(p+q-m-n-1)/2$,
$b_4=(m+n-p-q+3)/2$,
$b_5=(m-n-p-q+3)/2$.

This formula is relevant for identical fermions with a short range interaction 
in a quasi one-dimensional harmonic 
trap.
 
Corresponding results are shown in figure  2. 
Note that units are different from figure  1. 

Forward scattering is almost completely 
suppressed and the backward scattering interaction coefficients $V_{\rm c}$ become dominant. 
The $N$--dependence of $V_{\rm c}^{({\rm p})}$ ist different 
from that of $V^{(s)}$. We find
\begin{equation}\label{Exponentp}
V^{({\rm p})}_c \propto N^{0.5},
\end{equation}
which agrees for $L\approx L_{\rm F}$ completely with equation (\ref{A.11}) below in
the limit $k_{\rm F}\, d \ll 1$,
even if our interaction model in the appendix is extremely crude.
The dominance of $V_{\rm c}$ over $V_{\rm a}$ and $V_{\rm b}$ persists for all $N$. The momentum non-conserving coefficients also
increase with $N$ but slightly less than $V_{\rm c}$.

As was discussed in Section 2, the Fermi algebra removes 
any fully symmetric part of the interaction coefficients $V(m,p;q,n)$.
Inspection of figures 1 and 2 shows that not only the two 
forward scattering peaks 
disap\-pear after this procedure but also almost all momentum non-conserving
coefficients. In principle, we considered an arbitrary 
finite ranged potential $V(z)$ here but
the approximation (\ref{Gen5}) and (\ref{2.4}) require an interaction of short
range to be valid.
Note also that figures 3 and 4 below show that the same feature is present
in the case of the marginally long ranged dipole-dipole interaction.
We conclude that theories like \cite{WW01_2} 
which exactly solve the model
find strongest support by our results if
they take into account backscattering between identical fermions.

\section{Dipole--Dipole Interaction}

The real space interaction between two parallel magnetic dipoles $\mbox{\boldmath$\mu$}$ is

\begin{eqnarray}\label{4.1}
V_{\rm 3DD}({\bf x}) = \frac{\mu _0}{4 \pi} \left[ \frac{\mu^2}{r^3}-3 
\frac{ ({\bf x}\cdot \mbox{\boldmath$\mu$})^2 }
{r^5}-\frac{8 \pi}{3} \mu^2 \delta^{(3)}({\bf x})\right].
\end{eqnarray}

The Fourier transform becomes \cite{GER01}

\begin{eqnarray}\label{4.2}
\tilde{V}_{\rm 3DD}({\bf k}) =  \mu _0 \left(
\frac{(\mbox{\boldmath$\mu$} \cdot {\bf k})^2}{k^2}-\mu^2\right).
\end{eqnarray}

These functions are singular at ${\bf x}={\bf 0}$ and ${\bf k}={\bf 0}$, respectively.

With trivial redefinitions, all formulae can be taken over to the case of molecules
interacting via permanent electric dipoles.

The Fourier version of equation (\ref{1.7a}) reads

\begin{eqnarray}\label{4.4}
\tilde{V}_{\rm 1eff}(k_\|) = \int \frac{\d^2k_\perp}{(2 \pi)^2} 
\exp\left(-k_\perp^2/2 \alpha^2_\perp\right) \,\tilde{V}_3 ({\bf k}),
\end{eqnarray}

where $k_\|$, $k_\perp$ are the components of the wave vector $\bf k$
with respect to the axis of the trap.

A straightforward application of equation (\ref{4.4}) gives the effective one-dimensional dipole-dipole 
potential

\begin{eqnarray}\label{4.4a}
\fl \tilde{V}_{\rm 1eff\,DD}(k_\|) = - \frac{\mu _0 \mu^2 \alpha _\perp^2}{2 \pi}\,
\left\{\frac{3\cos^2\Phi-1}{2}\left[1+\frac{\kappa^2}{2}\,e^{\kappa^2/2}\,{\rm Ei}(-\kappa^2/2) \right]
+\sin^2 \Phi \right\}
\end{eqnarray}

with $\kappa=k_\|/\alpha_\perp$. 
$\rm Ei$ 
denotes
the exponential integral 
and $\Phi$ is
the angle between the dipoles and the axis of the trap.

The effective one-dimensional potential is non-analytic at $k=0$, but has the finite value

\begin{eqnarray}\label{4.6}
\tilde{V}_{\rm 1eff\,DD}(k_\|=0) = 
- \frac{\mu _0 \mu^2 \alpha _\perp^2}{4 \pi}(1+\cos^2 \Phi).
\end{eqnarray}

For identical fermions, this value must be subtracted from equation (\ref{4.4a}), as 
was explained above.

It is possible to calculate exactly the real space version of the effective one-dimensional dipole--dipole 
interaction:

\begin{eqnarray}\label{4.7}
\fl V_{\rm 1eff\,DD}(z) =\frac{1}{2\, \pi}\,\int _{-\infty}
^\infty \d k_\|\,e^{-i k_\| z}\,\tilde{V}_{\rm 1eff\,DD}(k_\|) 
\\[4mm]\nonumber
\lo=   - \frac{\mu _0 \mu^2 \alpha _\perp^3}{8\sqrt{2 \pi}}
\bigg\{(3\cos^2\Phi-1) \bigg[(1+\alpha_\perp^2 z^2)\,
\exp\left(\alpha_\perp^2 z^2/2\right)\,{\rm erfc}\left(\alpha_\perp |z|/\sqrt{2}\right)\\[4mm]\nonumber
- \sqrt{2/\pi}\,\alpha _\perp |z|\bigg] 
  + \frac{8}{\sqrt{2\pi}\,\alpha_\perp} \sin^2 \Phi \,\, \delta(z)\bigg\}.   
\end{eqnarray}

For $|\Phi|< (>) \arccos\, 1/3$, the potential is attractive (repulsive).
Note that in a one-di\-men\-sio\-nal Fermi gas quantum fluctuations prevent the classical 
collapse of this configuration
until a threshold attractive interaction
strength is reached which according to \cite{WW01_2} is
$|V_{\rm a,b,c}|=O(\hbar \omega_\ell)$.
The range $d_{\rm 1D}$ of this effective one-dimensional potential inside the trap follows from 
$d_{\rm 1D}\alpha_\perp=1$, which gives $k_{\rm F} d_{\rm 1D}=\sqrt{2}$ for a completely filled trap. 
The real space potential, which has a cusp at $z=0$, is shown in the inset of figure  4 for the case of dipoles oriented parallel to the axis of the trap
($\Phi=0$).

The large distance behaviour is 

\begin{eqnarray}\label{4.8}
V_{\rm 1eff\,DD}(|z| \gg \alpha _\perp^{-1}) \rightarrow  - \frac{\mu _0 \mu^2 }{4 \pi}\,(3\cos^2 \Phi-1) 
\, \frac{1}{|z|^3}.            
\end{eqnarray}

This is exactly the behaviour of the initial potential equation in three dimensions (\ref{4.1}) along the 
$z$-axis for two parallel dipoles.    

We now turn to the interaction coefficients for the dipole--dipole interaction.
The analytic procedures outlined in equations (\ref{Gen2}) to (\ref{2.4}) are not applicable to the dipole--dipole
interaction. We thus resort to a numerical evaluation and consider
the case of dipoles oriented parallel to the $z$--axis:
The sequence $V(m=107,p=103;q=105,n)$ of interaction coefficients is displayed in figure  3, as
calculated from the full equation (\ref{4.4}), i.e., for the interaction between fermionic
atoms in different hyperfine states. 
Here and in figure 4 a filling factor 
$F=N\omega_\ell/\omega_\perp \approx 1$ is assumed.

It is seen that the backward scattering coefficient
$V_{\rm c}$ is suppressed in comparison to $V_{\rm a} \approx V_{\rm b}$ (for large $N$, $V_{\rm a} \rightarrow
V_{\rm b}$ hold due to the long range of the interaction. This is in accordance with the results for the model interaction in the box 
(equation (\ref{2.7}) below, case $2 k_{\rm F} d=O(1))$.

The $N$--dependence of the coefficients $V_{\rm a}$, $V_{\rm b}$, and $V_{\rm c}$ 
for the full dipole-dipole interaction is
\begin{equation}
V_{\rm a}, V_{\rm b} \propto N^{-0.3}, \qquad V_{\rm c} \propto N^{-0.1}
\end{equation}
while the background coefficients which dominate more
than in the case of the contact interaction (cf. figure 1)
remain almost constant.

In figure  4, we show the interaction coefficients for identical fermions, i.e.,
for the effective one-dimensional potential $\tilde{V}_{\rm 1eff\,DD}(k)-\tilde{V}_{\rm 1eff\,DD}
(k=0)$. It can be seen that in this case the approximation

\begin{equation}\label{4.9}
V(m,p;q,n) \approx V_{\rm c}^{({\rm non-s})}\,\,\delta _{m+q,n+p}
\end{equation}

is appropriate, i.e., only backward scattering is relevant.

The $N$--dependence of $V_{\rm c}^{\rm \rm (non-s)}$ is different
for different filling factors
 $F=N \omega _\ell/\omega _\perp$. 
For $F\approx 1$ the $N$--dependence is estimated as 
$V_{\rm c}^{\rm (non-s)} \propto N^{-0.5}$ while for $F\ll 1$ it becomes
$V_{\rm c}^{\rm (non-s)} \propto N^{0.5}$.
Both results
agree with equation ({\ref{Exponentp}) as well as with equation (\ref{A.7})
(note that in the first case $k_{\rm F}\, d \approx 1$
while $k_{\rm F}\, d \ll 1$ for the second case).
The background coefficients grow slightly with $N$.

From our results, we conclude that neglect of the background
coefficients finds justification for
the case of identical fermions in traps with
a small filling factor after the s-wave part has been subtracted.

Finally, we estimate $V_{\rm a} \approx V_{\rm b}$ for fermions in different hyperfine states,
which interact via the magnetic dipole--dipole interaction.
For $m, p, q, n \approx N$ the coefficients
$V^{({\rm s})}_{\rm a,b,c}(m,p;q,n)$ do not exceed
$V(N,N;N,N)$. With dipoles
oriented parallel to the trap axis and using the ''s-wave'' approximation
in equation (\ref{2.7}) we get

\[
V^{({\rm s})}(N,N,N,N) \approx -1.34 \,\frac{\mu_0 \mu^2}{4\pi}\,\frac{m_{\rm A}}{\hbar^2}\,\frac{\alpha_\perp}{\sqrt{F}}\,\,\hbar\omega_\ell.
\]

We refer to the planned highly an\-iso\-tro\-pic trap in \cite{Ott01}
and find with $\omega_\perp=12\pi \cdot 10^5 s^{-1}$ for   
 $^{53}{\rm Cr}$:

$$\left|\frac{V^{({\rm s})}_{\rm a,b,c}}{\hbar \omega_\ell}\right| < \frac{0.2 }{\sqrt{F}}.$$

Provided $F$ is small, $|V_{\rm b}|$ can become of the order of the unperturbed level
splitting $\hbar \omega _\ell$, which is required for interactions to be relevant.

\section{Discussion and Summary}

The prospect of realizing a quasi one-dimensional gas of
interacting fermions in a trap will allow
a comparison of experiments with emerging theories
\cite{WW01_2,Recati} based on Luttinger liquid theory. 
These theories require a knowledge of the coupling constants 
describing the interaction.

We have shown in detail, how a physical pair interaction between
fermions in three spatial dimension determines three effective coupling
functions $V_{\rm a}$, $V_{\rm b}$, and $V_{\rm c}$ for the interaction of the fermions
when they are confined to a quasi one-dimensional harmonic trap.

The coupling functions $V_{\rm a}$ and $V_{\rm b}$ describe the forward
scattering, while $V_{\rm c}$ quantifies the backward scattering. Their 
respective values depend significantly on the range $d_{\rm 3D}$ of the 
original pair interaction, the reference length being the 
inverse of the one-dimensional
Fermi wave number. 

For $2k_{\rm F} d_{\rm 3D} \ll 1$ and distinguishable fermions, i.e., for the inter-component 
interaction in a quantum gas mixture,
e.g. for the van der Waals interaction, 
forward and backward scattering contribute
about equally. The effective one-dimensional potential is also short ranged and the equivalent
of the s-wave approximation is applicable.

For dipole--dipole interaction where $d_{\rm 3D}$ is not defined,
the effective one-dimensional potential
acquires a range $d_{\rm 1D} \approx 1/k_{\rm F}$ inside the trap.

For distinguishable fermions,
backward scattering becomes suppressed in comparison 
with forward scattering 
with increasing range of interaction.

These results change when the fermions are identical. The contact potential ("s-wave
scattering") then does not contribute. This has a surprising effect on the effective
coupling functions: To a good approximation, only backscattering survives, i.e.,
$V_{\rm a}$ and $V_{\rm b}$ can be neglected in comparison with $V_{\rm c}$ 
even if this  backscattering coefficient is
significantly reduced in the case of 
short range interactions. 

For the dipole-dipole interaction, however, it is of the same
order as for the full dipole--dipole interaction.
In reality, only the electric dipole--dipole interaction
between identical fermionic molecules with a permanent 
electric dipole moment \cite{BBC00} 
can become relevant.  
Finally we mention that dipolar fermionic molecules
$^{40}\rm K$-$^{87}\rm Rb$ were discussed \cite{Roati02} as possible 
candidates for fermionic superfluidity.

\ack
We gratefully acknowledge helpful discussions with G. Alber, X. Gao, T. Pfau, 
and W. P. Schleich and financial help by Deutsche Forschungsgemeinschaft. 

\Figures
\begin{figure}
\caption{Reduced interaction coefficient $V^{({\rm s})}(1220,1214;1200,n)
/(\alpha \langle V \rangle)$ for ''s-wave''
scattering in the quasi one-dimensional harmonic trap. Three contributions at $n=1194$
 ($V_{\rm b}$), $n=1206$ ($V_{\rm c}$), and $n=1234$ ($V_{\rm a}$) dominate.
Coefficients with an odd sum
of arguments vanish.}
\end{figure}

\begin{figure}
\caption{Reduced interaction coefficients 
$V^{({\rm p})}(1220,1214;1200,n)/(\alpha^3 \frac{1}{2}\langle z^2\, V\rangle )$ 
for ''p-wave'' scattering
versus $n$ for the quasi one-dimensional harmonic trap. Backscattering coefficient 
$V^{({\rm p})}_c < 0$ at $n=1206$ dominates.
Coefficients with an odd sum
of arguments vanish.
Background coefficients as seen in figure  1 almost disappeared.}
\end{figure}

\begin{figure}
\caption{Reduced interaction coefficients $V(107,103;105,n)/|{\tilde V}_{\rm 1eff\,DD}(0)|$ for the 
full dipole--dipole 
interaction between atoms in different hyperfine states versus number $n$ of oscillator state. 
Dipoles are oriented parallel to the trap axis and a filling factor of $\approx$ 1 is assumed.
Backscattering coefficient $V_{\rm c}$ at $n=109$ is smaller than the forward scattering
coefficients $V_{\rm a}$ and $V_{\rm b}$, which also dominate the other interaction coefficients.
Coefficients with an odd sum
of arguments vanish.}
\end{figure}

\begin{figure}
\caption{Reduced interaction coefficients $V^{\rm (non-s)} (107,103;105,n)/|{\tilde V}_{\rm 1eff\,DD}(0)|$ for the 
dipole--dipole 
interaction between identical atoms versus number $n$ of oscillator state. 
Dipoles are oriented parallel to the trap axis and a filling factor of $\approx$ 1 is assumed.
Backscattering coefficient $V^{\rm (non-s)}_c$ at $n=109$ dominates all other interaction 
coefficients including the forward scattering coefficients. 
Moreover, background coefficients are much smaller in comparison
with the main peak than in figure  3.
The inset shows the real 
space version of the effective one-dimensional potential $V_{\rm 1eff\,DD}$ versus $z$.
Coefficients with an odd sum
of arguments vanish.}
\end{figure}

\appendix
\section*{Appendix: Model interaction in the box}
 
\renewcommand{\theequation}{A.\arabic{equation}}
Equations (\ref{2.7}) and (\ref{2.8}) allow easy numerical evaluation, but are not 
transparent. We, therefore, study a solvable model for the interaction coefficients.
We consider the one-dimensional potential (cf. also \cite{SchMM00} for a related analysis) 

\begin{equation}\label{A.1}
V(z)=\frac{g}{2d}\,\,\,\, {\rm for}\,\, |z|\le d \,\,\,{\rm and }\,\,\, 
V(z)=0 \,\,\, {\rm for}\,\, |z|>d,
\end{equation}

normalized according to
\[\int _{-\infty}^\infty \d z\, V(z)=g.\]

The fermions are confined to a box of length $L \gg d$ with infinitely high walls. The single 
particle eigenstates are then

\begin{equation}\label{A.2}
\psi _m(z)=\sqrt{\frac{2}{L}}\,\sin \frac{m\pi z}{L},\quad m=1,2,3,\ldots\,.
\end{equation}

The interaction coefficients according to equation (\ref{1.5}) are given (up to a small 
correction of order $d$) by 

\begin{eqnarray}\label{A.4}
\fl V_{\rm box}(m,p;q,n;d)
  =  \frac{2g}{d L^2}\int _0^{L} \d z_1 
\sin \frac{m\pi z_1}{L}
\sin \frac{q\pi z_1}{L}
\int _{z_1-d}^{z_1+d} \d z_2\,
\sin \frac{p\pi z_2}{L}
\sin \frac{n\pi z_2}{L}
\\[4mm]\nonumber
  =  \frac{1}{2L}\left(\frac{\sin (p-n)\pi d/L}{(p-n)\pi d /L}
(\delta _{m+p,q+n}+\delta _{m+n,p+q})+
 \frac{\sin (p+n)\pi d/L}{(p+n)\pi d/L}
\delta _{m+q,p+n}\right).
\end{eqnarray}

All coefficients, except those with
\begin{equation}\label{A.3}
m \pm p \pm q \pm n = 0,
\end{equation}
vanish due to local translational invariance inside the box.
But near the Fermi surface ($m,p,q,n \approx N \gg 1$)
only the three processes shown in equation  (\ref{A.4}) remain: 
two with small and one with large momentum transfer $\Delta p$ as was discussed in Section 2.
 
In the notation of equation (\ref{1.6}), this can be written as

\begin{equation}\label{A.5}
V_{\rm box}(m,p;q,n;d)=V_{\rm a}\,\delta _{m+p,q+n}+V_{\rm b}\,\delta _{m+n,p+q}
+V_{\rm c}\,\delta _{m+q,p+n}.
\end{equation}

Again, for processes near the Fermi edge and $d\ll L$ we approximate

\begin{equation}\label{A.6}
V_{\rm a} = V_{\rm b} \approx \frac{g}{2L}\left( 1- \frac{(p-n)^2 \pi^2 d^2}{6 L^2}\right),
\end{equation}

\begin{equation}\label{A.7}
V_{\rm c} \approx \frac{g}{2L}\left( \frac{\sin( 2N\pi d/L)}{2N\pi d/L}\right).
\end{equation}

The coefficients depend
on the
differences between the indices only ($V_{\rm a,b}$) or are constant with respect
to the indices ($V_{\rm c}$). 
We note in passing that
the interaction coefficients $V_{\rm box}(m,p;q,n;d)$ show perfectly the behaviour
which allows application of the Bosonization method, see \cite{WW01_2}.

In the limit $d\rightarrow 0$
the potential (\ref{A.1}) becomes a contact potential $g\,\delta(z)$
with pure $s$-wave scattering
and all three types of coefficients
are equal and constant

\begin{equation}\label{A.8}
V_{\rm a,b,c} \rightarrow V_{\rm a,b,c}^{({\rm s})} = \frac{g}{2L}.
\end{equation}

However, when the range $d$ of the potential (\ref{A.1}) 
increases and
becomes larger than the average distance $L/N$ of the atoms in the box 
the backscattering coefficient $V_{\rm c}$ decreases while $V_{\rm a,b}$
remain constant as long as $d\ll L$.

As discussed in Section 3 above, 
any fully symmetric part common to all interaction coefficients
does not contribute to
the interaction ${\hat V}_3$ according to equation (\ref{1.1a}) and may therefore be removed.
Doing so we subtract the fully symmetric potential $V_{\rm box}(m,p;q,n;d=0)$
(\ref{A.8}) from $V_{\rm box}(m,p;q,n;d)$ (\ref{A.5},\ref{A.6},\ref{A.7}) with the result that only a 
backscattering term which we denote by $V_{\rm c}^{({\rm p})}$
remains 

\begin{equation}\label{A.9}
V_{\rm box}(m,p;q,n;d)  \rightarrow  V_{\rm c}^{({\rm p})}\, \delta_{m+q,p+n}
\end{equation}
\begin{eqnarray}\label{A.10}
V_{\rm c}^{({\rm p})}  & = & \frac{g}{2L} \left(\frac{\sin (2\pi N d/L)}{2\pi N d/L}-1\right)\, \delta_{m+q,p+n}
\\[4mm]\nonumber
 &\approx & - \frac{g}{3L} \, k_{\rm F}^2d^2\, \delta_{m+q,p+n},
\end{eqnarray}

where the latter approximation is valid for 
$k_{\rm F} d \ll 1 $ ($k_{\rm F}=\pi N/L$).
The sign of $V_{\rm c}^{({\rm p})}$ is opposite to the sign of 
$V_{\rm c}^{({\rm s})}$ in accordance with the results for the harmonic trap, compare
figures 1 and 2.

In equations  (\ref{Exponent}) and
(\ref{Exponentp}) we found in the case of the harmonic trap
for the scaling of the coefficients
with the particle number $N$: $V_{\rm a,b,c}^{({\rm s})}\sim N^{-0.3}$ and
$V_{\rm c}^{({\rm p})}\sim N^{0.5}$. To compare with equations  (\ref{A.8})
and (\ref{A.9}) where the coefficients are given in terms of
$L$ and $k_{\rm F}$ we recall that in the harmonic trap 
$k_{\rm F}$ and $L$ are proportional to $\sqrt{2N}$.
We substitute these relations into the box coefficients and get
\begin{equation}\label{A.11} 
V_{\rm a,b,c}^{({\rm s})} \sim N^{-1/2},\qquad V_{\rm c}^{({\rm p})}\sim N^{1/2}.
\end{equation} 
We conclude that in spite of the crude character of our model
the agreement of the cofficients $V_{\rm a,b,c}$ in the box and in
the harmonic trap is quite good. The background terms for the
momentum non-conserving processes in the harmonic trap are,
of course, absent in the box.


\begin{thebibliography}{99}
\bibitem{model} 
One must expect that a renormalization of the bare values
occurs. For practical applications, this may not be a serious problem because only the central
coupling constants which describe the interaction in the mass and composition
channels are relevant. As in the Luttinger model these are treated
as adjustable parameters.
\end{thebibliography}
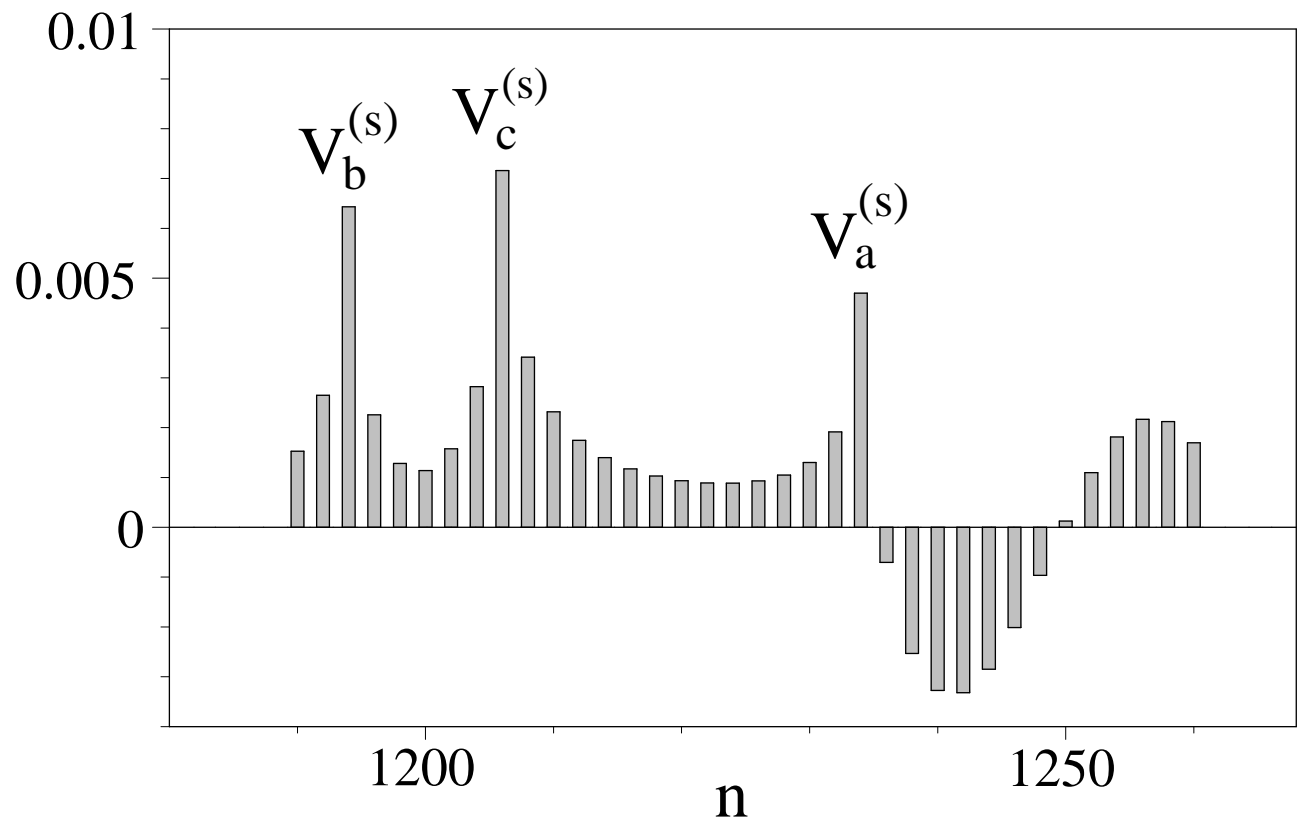
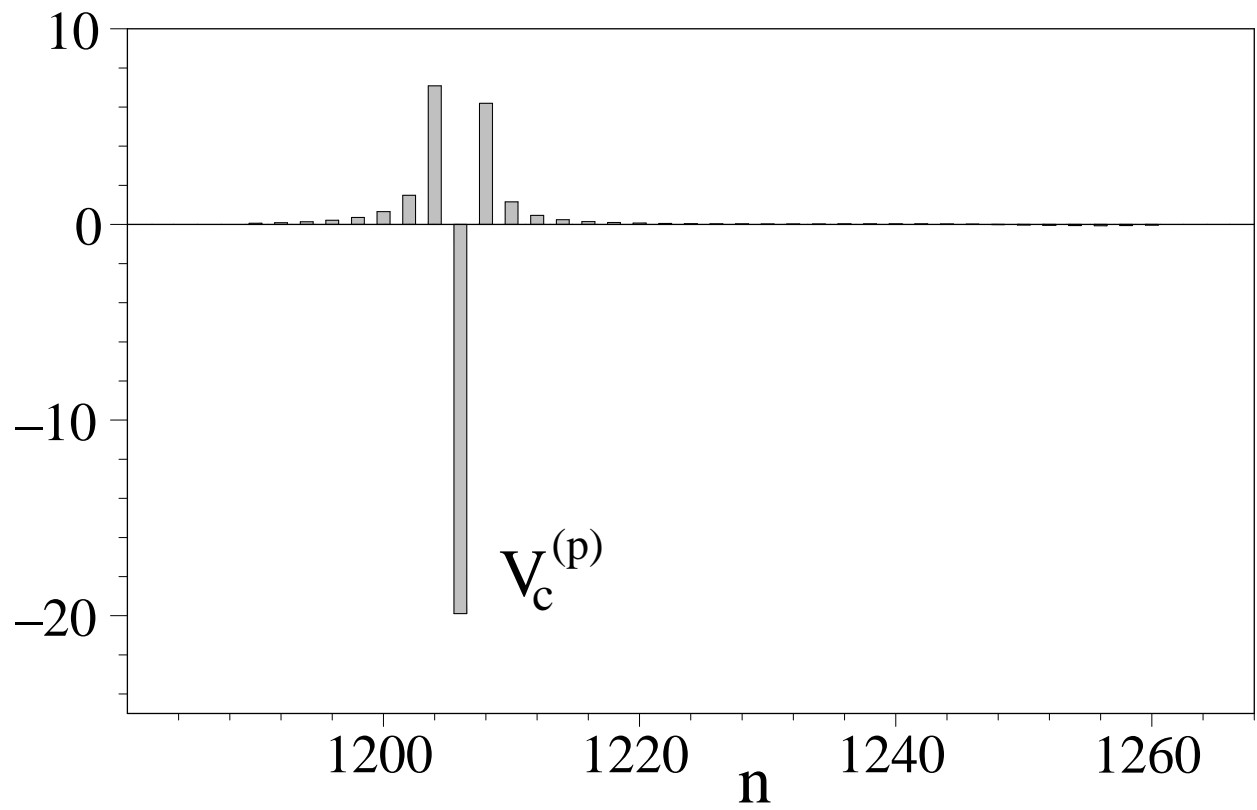
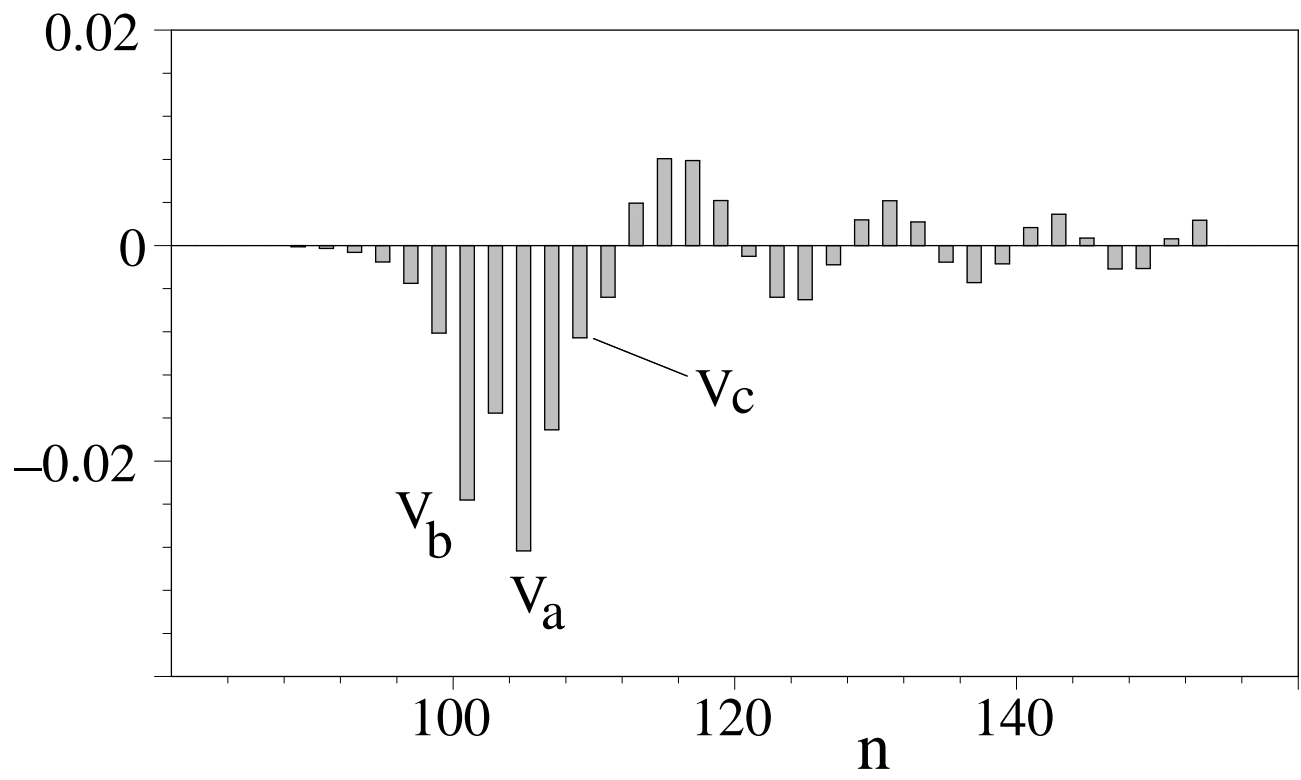
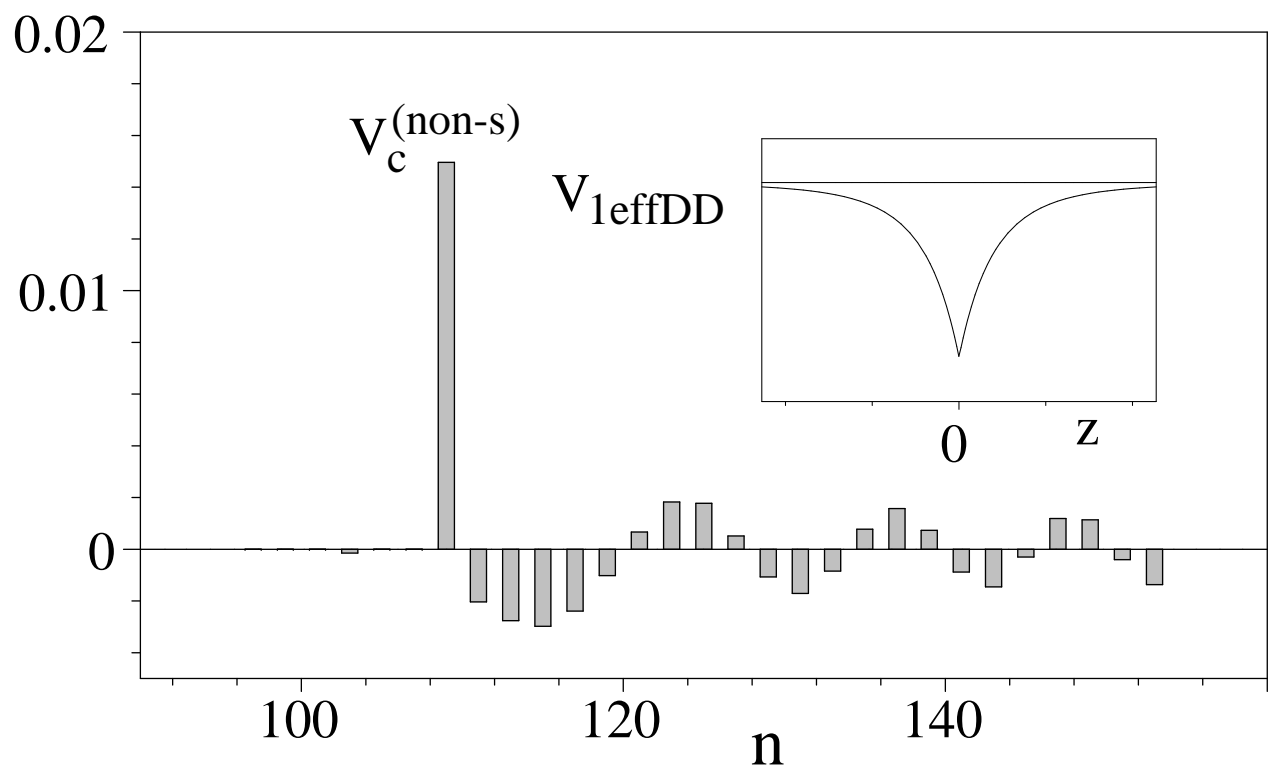
\end{document}